\documentclass[
  aps,
  prc,
  reprint,
  superscriptaddress,
  amsmath,amssymb,
  nofootinbib
]{revtex4-2}

\usepackage{graphicx}
\usepackage{dcolumn}
\usepackage{bm}
\usepackage{hyperref}
\usepackage{xcolor}

\begin{document}

\title{Breakdown of the Plane-Wave Trojan Horse Analysis of the
$^{12}\mathrm{C}+{}^{12}\mathrm{C}$ Fusion Reaction: Critical Role of Coulomb Distortions}

\author{A. M. Mukhamedzhanov}
\affiliation{Texas A\&M University, College Station, Texas 77843, USA}

\date{\today}

\begin{abstract}
Recently, a new Trojan Horse Method (THM) measurement of carbon-carbon fusion
was reported by Li \textit{et al.} [Phys. Lett. B (2026) 140675].
The purpose of the present work is to demonstrate the breakdown of the
plane-wave approximation used in the analysis of these data and the critical
role of Coulomb distortions in the initial and final states.

The reaction mechanism underlying the THM analysis of the
$^{12}\mathrm{C}+{}^{12}\mathrm{C}$ fusion reaction using the
$^{16}\mathrm{O}+{}^{12}\mathrm{C}\to
\alpha_s+\alpha+{}^{20}\mathrm{Ne}$ reaction is investigated.
Particular attention is paid to the spectator momentum distribution and to
the dependence of the THM reaction amplitude on the relative carbon-carbon energy
$E$.

It is demonstrated that agreement with the measured spectator momentum
distribution does not by itself validate the plane-wave approximation.
Although the experimental momentum distribution can be reproduced,
inclusion of Coulomb distortions in both the initial and final channels
leads to an energy dependence of the THM amplitude that is completely
different from the plane-wave result. Consequently, the energy dependence
of the $^{12}\mathrm{C}+{}^{12}\mathrm{C}$ fusion cross section extracted
from the THM data can be strongly distorted by the plane-wave treatment.
It is concluded that the astrophysical factor extracted in the
plane-wave analysis cannot be regarded as reliable and may lead to
misleading conclusions concerning the low-energy
$^{12}\mathrm{C}+{}^{12}\mathrm{C}$ fusion reaction.
\end{abstract}

\maketitle

\section{Introduction}
\label{sec:intro}

The low-energy $^{12}\mathrm{C}+{}^{12}\mathrm{C}$ fusion reaction plays an
important role in carbon burning and in a variety of stellar environments.
Direct measurements at astrophysically relevant energies are extremely
difficult because of the strong Coulomb suppression. The only direct
measurement extending to relative carbon-carbon energies $E<2.2$ MeV,
performed in inverse kinematics, was reported in Ref.~\cite{NanWang}.
This measurement represented an important breakthrough in the study of
carbon-carbon fusion.

Before the direct measurement of Ref.~\cite{NanWang}, the only
experimental access to energies below $E\simeq 2.1$ MeV was provided by
the Trojan Horse Method (THM) measurement reported in Ref.~\cite{Tumino}. A major shortcoming of
that analysis, however, was the use of the plane-wave approximation (PWA),
which led to a sharp increase of the extracted astrophysical $S$ factor
toward low energies. Such an increase was not confirmed by the subsequent
direct measurement of Ref.~\cite{NanWang}.

Recently, a new THM measurement was reported in Ref.~\cite{Li}, employing
the three-body reaction
\begin{align}
{}^{12}{\rm C}({}^{16}{\rm O},\alpha_s\alpha){}^{20}{\rm Ne},
\label{16O12C}
\end{align}
where the incident Trojan-horse nucleus is regarded as
${}^{16}{\rm O}={}^{12}{\rm C}+\alpha_s$ and $\alpha_s$ is the spectator.
As in Ref.~\cite{Tumino}, the analysis of Ref.~\cite{Li} relies on the PWA.

The goal of the present work is to demonstrate the breakdown of the PWA
used in Ref.~\cite{Li} and to investigate its consequences for the
extracted energy dependence of the THM reaction amplitude. The use of the
PWA in Ref.~\cite{Li} was supported by the observation that the measured
spectator momentum distribution can be reproduced by the momentum
distribution obtained from the Fourier transform of the
${}^{16}{\rm O}={}^{12}{\rm C}+\alpha$ bound-state wave function generated
with a Woods--Saxon potential. It is shown that agreement with the measured
spectator momentum distribution does not constitute sufficient
justification for the PWA\@. In particular, such agreement does not imply
that Coulomb distortions in the initial and final channels can be neglected.

This issue is especially important for the present heavy-ion reaction,
where the Coulomb interactions in both the entrance and exit channels are
strong. The formalism of Ref.~\cite{TypelBaur}, invoked in support of the
analysis of Ref.~\cite{Li}, was developed and illustrated for THM
applications involving considerably lighter systems. Its approximations
therefore cannot be assumed, without further justification, to remain
valid for the strongly Coulomb-distorted
${}^{16}{\rm O}+{}^{12}{\rm C}$ reaction considered here. It is demonstrated
explicitly that inclusion of Coulomb distortions can leave the spectator
momentum distribution rather weakly modified while producing a
qualitatively different dependence of the THM amplitude on the
$^{12}{\rm C}+{}^{12}{\rm C}$ relative energy.

Despite the use of the same PWA as in Ref.~\cite{Tumino}, the extracted
overall trend of the $S$ factor does not reproduce the sharp rise toward low
energies that was a characteristic feature of the result reported in
Ref.~\cite{Tumino}. While the THM measurement of Ref.~\cite{Tumino} and the
direct measurement of Ref.~\cite{NanWang} exhibit a rich resonance structure,
the $S$ factor extracted in Ref.~\cite{Li} instead displays relatively broad
oscillatory structures, with individual resonances poorly resolved.
This difference may be attributed, at least in part, to the limited energy
resolution of the experiment in Ref.~\cite{Li}.

The reaction amplitude itself is examined, with particular emphasis on the
Coulomb interactions in both the entrance and exit channels. 
In the present work, attention is restricted to the Coulomb distortions in the initial and final states. This restriction is particularly important for the THM analysis because of the long-range character of the Coulomb interaction, associated with the zero mass of the photon. In contrast to short-range nuclear distortions, these Coulomb interactions modify the magnitude and energy dependence of the reaction amplitude without shifting the quasi-free peak or the resonance-pole positions of the intermediate subsystem. They can therefore produce a substantial renormalization of the extracted THM astrophysical factor while preserving the locations of the physical resonances. It is shown that
the spectator momentum distribution and the energy dependence of the THM
amplitude provide distinct tests of the reaction mechanism.

\section{Short introduction to THM}
\label{IntTHM}

The THM can be used to infer the low-energy behavior of the $S$ factor
for a binary reaction from a suitable three-body surrogate process in which
the binary reaction of interest appears as a subreaction. The principal advantage of
the THM is that it allows the astrophysical $S$ factor to be determined in
the Gamow-window region, which may lie deep below the Coulomb barrier, where
direct experimental data are often unavailable. However, extraction of the
$S$ factor from THM data is not straightforward and is model dependent,
because the required information must be inferred from a surrogate reaction
involving three particles in the final state. The analysis becomes
considerably more complicated when the three-body Coulomb interaction is
important.

\begin{figure}[t]
    \centering
    \includegraphics[width=0.95\columnwidth]{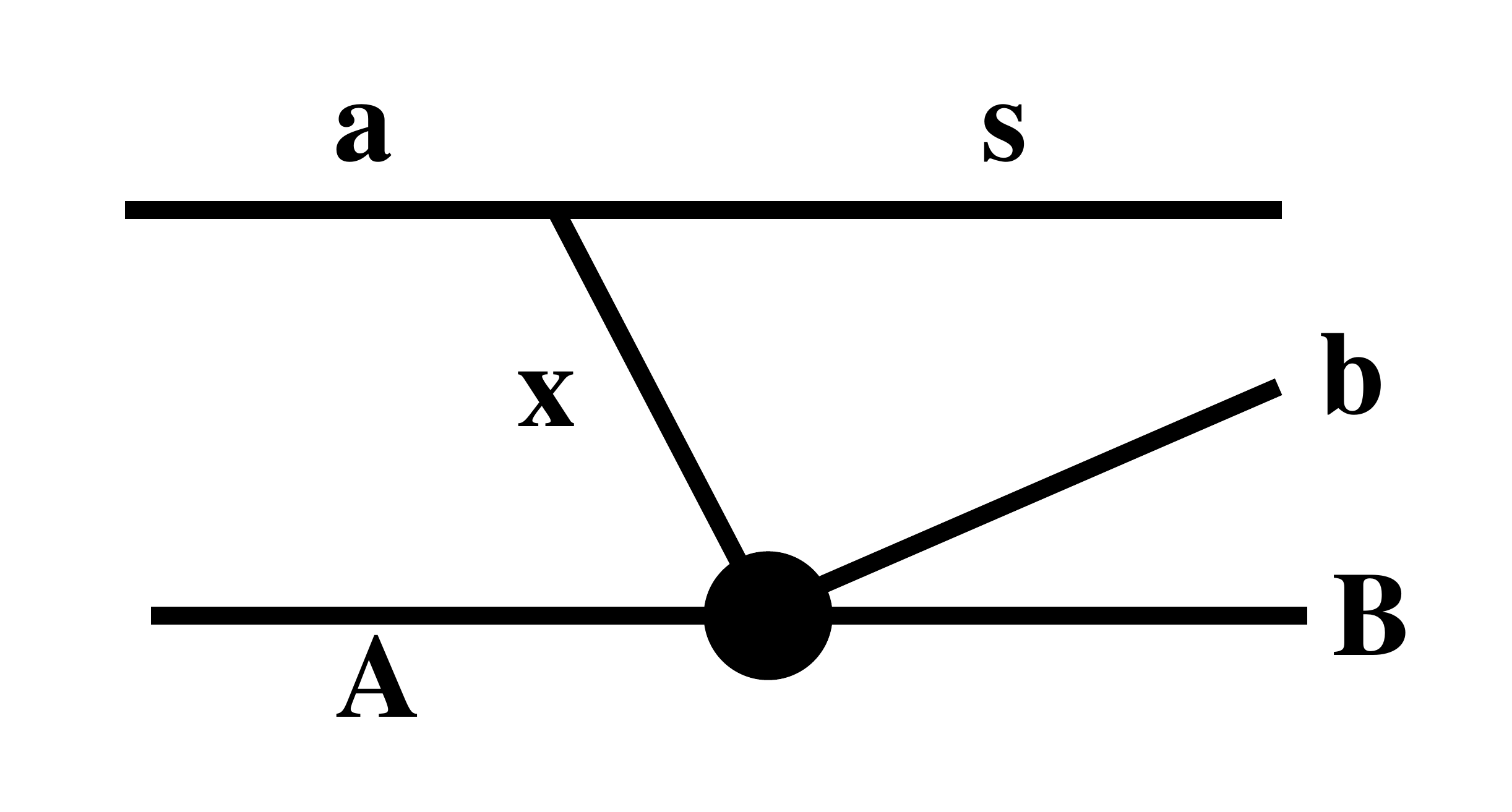}
    \caption{Skeleton diagram of a generic THM reaction
    $a+A\to s+b+B$, in which the virtual transferred particle $x$
    interacts with $A$ through the binary subreaction $x+A\to b+B$,
    while $s$ acts as the spectator. Initial- and final-state
    rescattering distortions are omitted.}
    \label{fig_PWA}
\end{figure}

Figure~\ref{fig_PWA} shows a skeleton diagram for a generic surrogate THM
reaction leading to the three-body final state $s+b+B$, with distortions due
to rescattering in the initial and final states omitted:
\begin{align}
a+A \to s+b+B.
\label{GenericTHM}
\end{align}

The goal of the THM is to extract the low-energy astrophysical factor,
or equivalently the cross section, of the binary subreaction
\begin{align}
x+A \to b+B
\label{subreaction1}
\end{align}
from measurements of the THM reaction~(\ref{GenericTHM}) performed at
energies for which the directly measured binary-reaction cross section
is strongly suppressed by the Coulomb and centrifugal penetrability
factor $P_{l_i}$. Here, $l_i$ is the orbital angular momentum in the
initial $x+A$ channel of the binary subreaction.

An essential feature of the THM mechanism is that the transferred
particle $x$ is virtual, that is, off the energy shell. This is advantageous
because the half-off-shell subreaction amplitude does not contain the usual
on-shell barrier penetrability factor in the entrance channel $x+A$. The
physical on-shell amplitude for the reaction $x+A\to b+B$ differs from its
half-off-shell counterpart, in which only the entrance-channel particle
$x$ is off the energy shell.

The subreaction~(\ref{subreaction1}) may correspond to elastic scattering,
a direct rearrangement reaction, or a resonance reaction. In general,
however, the THM cannot be used straightforwardly to determine the
amplitude of a direct transfer subreaction. The on-shell binary-reaction amplitude contains the penetrability factor $P_l$ in the
entrance channel $x+A$. Consequently, as $E\to0$, this factor strongly
suppresses all but the lowest partial waves, typically those with $l=0$
or $l=1$.

In the THM amplitude, by contrast, the absence of the corresponding
on-shell entrance-channel penetrability allows many partial waves to
contribute. Therefore, the partial-wave composition of the THM amplitude
may differ substantially from that of the physical on-shell binary
amplitude.

Only in special cases, for example when the $Q$ value of the direct
reaction is sufficiently large that the angular distribution becomes
nearly isotropic and is dominated by only a few low partial waves, may
the THM provide reliable information about the direct subreaction
amplitude.

\begin{figure}[t]
    \centering
    \includegraphics[width=1.0\columnwidth]{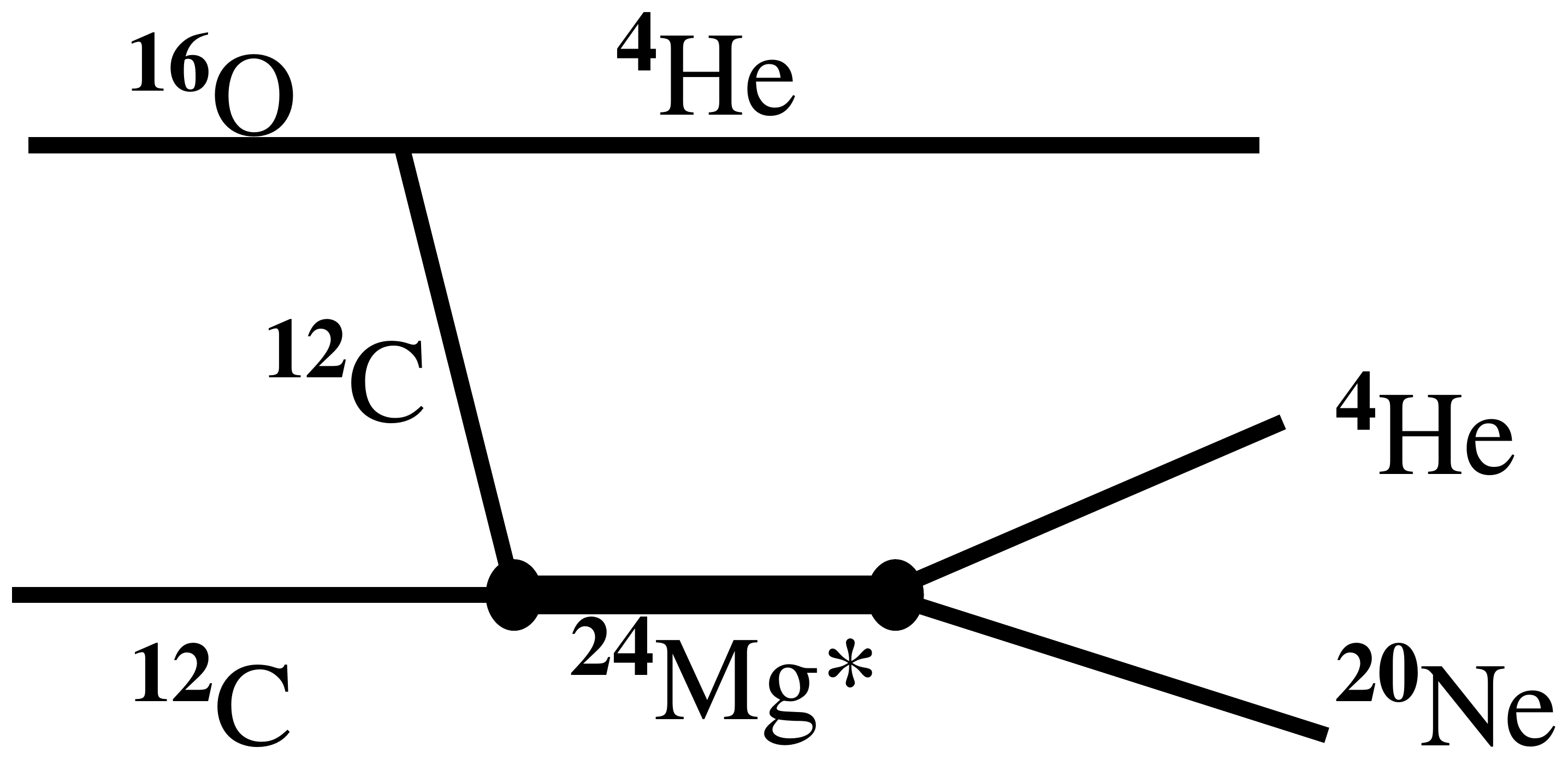}
    \caption{Skeleton Feynman diagram describing the THM reaction
    ${}^{16}{\rm O}+{}^{12}{\rm C}
    \to\alpha_s+{}^{24}{\rm Mg}^{*}
    \to\alpha_s+\alpha_0+{}^{20}{\rm Ne}$,
    which serves as a surrogate reaction for the binary resonant
    subprocess
    ${}^{12}{\rm C}+{}^{12}{\rm C}
    \to\alpha_0+{}^{20}{\rm Ne}$.}
    \label{fig_resPWA}
\end{figure}

The principal applications of the THM are resonant binary subreactions.
The corresponding skeleton diagram is shown in Fig.~\ref{fig_resPWA}.

\section{Kinematics of the THM experiment}
\label{sec:kinematics}

\subsection{Initial-state Coulomb effects in the
$^{12}{\rm C}({}^{16}{\rm O},\alpha\alpha_{0}){}^{20}{\rm Ne}$
Trojan-Horse reaction}

The reaction
\begin{equation}
^{16}\mathrm{O}+{}^{12}\mathrm{C}
\rightarrow \alpha_s+\alpha+{}^{20}\mathrm{Ne}
\label{eq:reaction}
\end{equation}
is considered \cite{Li}, where $^{16}\mathrm{O}=\alpha_s+{}^{12}\mathrm{C}$ and
$\alpha_s$ denotes the spectator particle.

A $28$ MeV ${}^{16}{\rm O}$ beam corresponds to a
${}^{16}{\rm O}-{}^{12}{\rm C}$ relative energy of $E_{i}=12$ MeV%
\footnote{The relative energy of two nuclei is invariant under a Galilean
transformation; that is, it does not depend on the coordinate system. In the
center-of-mass system (c.m.), it coincides with the c.m.\ energy of the
colliding nuclei.}.
The height of the Coulomb barrier of two colliding nuclei with atomic masses
$A_i$, $i=1,2$, is
\begin{align}
V_C =
\frac{Z_1Z_2e^2}
{r_0\left(A_1^{1/3}+A_2^{1/3}\right)}
\label{VCB1}
\end{align}
and the corresponding Coulomb parameter is
\[
\eta=
\frac{Z_1Z_2}{137}
\sqrt{\frac{\mu}{2E_{i}}},
\]
where the reduced mass $\mu$ of the ${}^{16}{\rm O}-{}^{12}{\rm C}$ system
and the energy $E_{i}$ are expressed in MeV.
For $r_0=1.4~{\rm fm}$ and the interaction radius
$R_{\rm ch}=1.4\left(16^{1/3}+12^{1/3}\right)\simeq6.73~{\rm fm}$,
the initial-state Coulomb barrier is $V_{CB}\approx10.3$ MeV and the
Coulomb parameter is $\eta_i=5.71$.

Thus, the initial-state relative energy exceeds the Coulomb barrier by only
$1.7$ MeV, while the corresponding Coulomb parameter remains large.

\subsection{Final-state Coulomb effects in the
$^{12}{\rm C}({}^{16}{\rm O},\alpha\alpha_{0}){}^{20}{\rm Ne}$
Trojan-Horse reaction}

After the virtual breakup
${}^{16}{\rm O}\rightarrow{}^{12}{\rm C}+\alpha_s$,
which precedes the formation of the intermediate
${}^{24}{\rm Mg}^{*}$ system, the total kinetic energy available to the
intermediate three-body system
$\alpha_s+{}^{12}{\rm C}+{}^{12}{\rm C}$ is
$E_{\rm int}=12.0-7.16=4.84~{\rm MeV}$, where
$\varepsilon=7.16$ MeV is the binding energy of the
${}^{16}{\rm O}=(\alpha\,{}^{12}{\rm C})$ bound state.
At
\[
E \equiv E_{{}^{12}{\rm C}-{}^{12}{\rm C}}=2.50~{\rm MeV},
\]
the spectator energy relative to the intermediate
${}^{24}{\rm Mg}^{*}$ subsystem is
\[
E_{\alpha_s-{}^{24}{\rm Mg}^{*}}
=
4.84-2.50
=
2.34~{\rm MeV}.
\]

For the $\alpha_0$ channel,
\[
E_{\alpha_0-{}^{20}{\rm Ne}}
=
2.50+4.62
=
7.12~{\rm MeV}.
\]

Using the Jacobi kinematics,
\[
E_{\alpha_s-\alpha_0}
=
3.66-2.88\cos\theta
\quad {\rm MeV},
\]
and
\[
E_{\alpha_s-{}^{20}{\rm Ne}}
=
2.47+1.34\cos\theta
\quad {\rm MeV}.
\]
Here, $\theta$ is the angle between the two Jacobi momenta,
\begin{align}
\cos\theta
&=
\frac{
\mathbf{k}_{\alpha_s-(\alpha_0{}^{20}\mathrm{Ne})}
\cdot
\mathbf{k}_{\alpha_0-{}^{20}\mathrm{Ne}}
}{
k_{\alpha_s-(\alpha_0{}^{20}\mathrm{Ne})}
k_{\alpha_0-{}^{20}\mathrm{Ne}}
}.
\end{align}
Thus, $\theta$ is the angle between the relative momentum
$\mathbf{k}_{\alpha_0-{}^{20}\mathrm{Ne}}$ in the decay
\begin{align}
{}^{24}\mathrm{Mg}^{*}
\to
\alpha_0+{}^{20}\mathrm{Ne},
\end{align}
and the momentum
$\mathbf{k}_{\alpha_s-(\alpha_0{}^{20}\mathrm{Ne})}$ of the spectator
$\alpha_s$ relative to the center of mass of the
$\alpha_0+{}^{20}\mathrm{Ne}$ pair.

The relative energies, Coulomb-barrier heights, and Coulomb parameters
for the relevant final-state subsystems are presented in
Table~\ref{tab_finstat}. The results show that the
$\alpha_s-\alpha_0$ subsystem changes from a subbarrier to an
above-barrier configuration as the Jacobi angle increases. Most
importantly, the spectator $\alpha_s$ is deeply subbarrier relative to the
intermediate ${}^{24}{\rm Mg}^{*}$ subsystem and remains subbarrier relative
to the final ${}^{20}{\rm Ne}$ nucleus for both representative angular
configurations. Thus, already at the lower boundary of the normalization
interval, $E=2.5$ MeV, where the THM $S^{*}(E)$ factor was normalized to the
Mazarakis \textit{et al.} data, the Coulomb effects in the final state are
sufficiently strong that they cannot be neglected.

Although the measurements were performed for both the
$\alpha_0+{}^{20}{\rm Ne}$ and
$\alpha_1+{}^{20}{\rm Ne}^{*}$ channels,
Table~\ref{tab_finstat} presents the kinematic conditions only for the
$\alpha_0+{}^{20}{\rm Ne}$ channel.
\begin{table*}[t]
\centering
\small
\renewcommand{\arraystretch}{1.35}
\caption{Relative energies, Coulomb barriers, and Coulomb parameters
for the relevant final-state subsystems. The identification of
$\theta=60^\circ$ with the laboratory detector opening angle is an
approximation; the exact Jacobi angle must be reconstructed event by event.}
\begin{tabular}{|l|c|c|c|c|l|}\hline
System & $\theta$ & $E_{\rm rel}$ (MeV) & $V_C$ (MeV) & $\eta$ & Coulomb regime \\
\hline
$\alpha_s-\alpha_0$ & $0^\circ$ & $0.78$ & $1.30$ & $1.01$ & subbarrier \\
\hline
$\alpha_s-\alpha_0$ & $60^\circ$ & $2.22$ & $1.30$ & $0.60$ & above barrier \\
\hline
$\alpha_s-{}^{20}{\rm Ne}$ & $0^\circ$ & $3.81$ & $4.78$ & $2.95$ & subbarrier \\
\hline
$\alpha_s-{}^{20}{\rm Ne}$ & $60^\circ$ & $3.14$ & $4.78$ & $3.25$ & more strongly subbarrier \\
\hline
$\alpha_s-{}^{24}{\rm Mg}^{*}$ & --- & $2.34$ & $5.52$ & $4.58$ & deeply subbarrier \\
\hline
\end{tabular}
\label{tab_finstat}
\end{table*}

The Coulomb effects become even stronger at the upper end of the
normalization interval. As the relative energy in the
${}^{12}{\rm C}+{}^{12}{\rm C}$ subsystem increases from
$2.5$ to $3.5$ MeV, the relative energy between the spectator
$\alpha_s$ and the intermediate ${}^{24}{\rm Mg}^{*}$ subsystem
decreases from $2.34$ to $1.34$ MeV. For the representative
configuration $\theta=60^\circ$, the
$\alpha_s-{}^{20}{\rm Ne}$ relative energy also decreases, from
approximately $3.14$ to $2.07$ MeV. Consequently, the final-state
Coulomb effects become even stronger toward the upper end of the
normalization interval.

\section{Spectator and internal momenta in the THM kinematics}
\label{sec:psx_psf}

It is important to distinguish the spectator Jacobi momentum
$\bm{k}_{sF}$ from the intrinsic momentum $\bm{p}_{sx}$ of the
$s+x$ configuration inside the Trojan-horse nucleus
$a=(sx)={}^{16}\mathrm{O}$. These momenta are related by
\begin{equation}
\bm p_{sx}=\bm k_{sF}
-\frac{m_s}{m_{sx}}\bm k_{aA}.
\label{eq:jacobi}
\end{equation}

For the reaction~(\ref{eq:reaction}) with
$a={}^{16}\mathrm{O}$, $s=\alpha_s$, $x={}^{12}\mathrm{C}$,
$A={}^{12}\mathrm{C}$, and $F=x+A={}^{24}\mathrm{Mg}^{*}$,
the two momenta are related by Eq.~(\ref{eq:jacobi}).
The corresponding relative momentum is
$k_{aA}\simeq391.5~\mathrm{MeV}/c$, and therefore
\[
\frac{m_s}{m_{sx}}k_{aA}\simeq97.9~\mathrm{MeV}/c.
\]

The total kinetic energy available in the intermediate
$\alpha_s+{}^{12}\mathrm{C}+{}^{12}\mathrm{C}$ system is
$E_{\rm int}=4.84~\mathrm{MeV}$.

Hence, for a given relative energy $E_{xA}$ in the
${}^{12}\mathrm{C}+{}^{12}\mathrm{C}$ subsystem,
\begin{equation}
E_{sF}=E_{\rm int}-E_{xA}=4.84-E_{xA}.
\label{eq:EsF}
\end{equation}

The spectator momentum is then
$k_{sF}=\sqrt{2\mu_{sF}E_{sF}}$.
Over the experimental interval
\[
0.5\leq E_{xA}\leq3.5~\mathrm{MeV},
\]
$E_{sF}$ varies over the interval $1.34$--$4.34~\mathrm{MeV}$.

Therefore,
$k_{sF}\simeq167~\mathrm{MeV}/c$ at $E_{xA}=0.5~\mathrm{MeV}$,
and
$k_{sF}\simeq92~\mathrm{MeV}/c$ at $E_{xA}=3.5~\mathrm{MeV}$.
Thus, the spectator Jacobi momentum covers approximately
\begin{equation}
\boxed{
92\lesssim k_{sF}\lesssim167~\mathrm{MeV}/c
}.
\label{eq:ksF_range}
\end{equation}

In the forward quasi-free configuration,
$\bm k_{sF}$ is approximately parallel to $\bm k_{aA}$,
so that Eq.~(\ref{eq:jacobi}) reduces to
\begin{equation}
p_{sx}
\simeq
\left|
k_{sF}
-
97.9
\right|.
\label{eq:psx_forward}
\end{equation}
At the upper end of the spectator-momentum interval,
$p_{sx}\simeq69~\mathrm{MeV}/c$,
whereas at the lower end,
$p_{sx}\simeq6~\mathrm{MeV}/c$.
Hence, the intrinsic momentum explored in the forward quasi-free
kinematics is only of order
\begin{equation}
\boxed{
6\lesssim p_{sx}\lesssim69~\mathrm{MeV}/c
}.
\label{eq:psx_range}
\end{equation}

More generally, if $\theta_{sF}$ is the angle between
$\bm k_{sF}$ and $\bm k_{aA}$,
\begin{equation}
p_{sx}^{\,2}
=
k_{sF}^{\,2}
+
(97.9~\mathrm{MeV}/c)^2
-
2k_{sF}(97.9~\mathrm{MeV}/c)
\cos\theta_{sF}.
\label{eq:psx_angle}
\end{equation}
For the experimental acceptance in the reaction center-of-mass system,
\begin{equation}
|\theta_{sF}|\leq18^\circ,
\end{equation}
the angular correction can be evaluated directly from
Eq.~(\ref{eq:psx_angle}). At the two ends of the experimental
$E_{xA}$ interval one obtains
\begin{align}
p_{sx}(k_{sF}=92~{\rm MeV}/c,\theta_{sF}=18^\circ)
&\simeq30.3~{\rm MeV}/c,\\
p_{sx}(k_{sF}=167~{\rm MeV}/c,\theta_{sF}=18^\circ)
&\simeq79.8~{\rm MeV}/c.
\end{align}
Consequently, even after allowing for the full c.m.\ angular acceptance,
the intrinsic momentum remains approximately bounded by
\begin{equation}
\boxed{p_{sx}\lesssim80~{\rm MeV}/c}.
\label{eq:psx_acceptance}
\end{equation}
This kinematic limit is substantially below the
$150$--$200~{\rm MeV}/c$ range displayed in Ref.~\cite{Li} in the
comparison between the experimental and calculated momentum distributions.

This discrepancy cannot be attributed to a choice of reference frame.
The intrinsic relative momentum $\bm p_{sx}$, defined by
Eq.~(\ref{eq:jacobi}), is invariant under a Galilean transformation and
therefore has the same value whether evaluated in the laboratory or in the
overall center-of-mass system. Thus, if the momentum denoted by $p_s$ in
Ref.~\cite{Li} is the intrinsic $\alpha_s$--${}^{12}{\rm C}$ momentum, its
extension to $150$--$200~{\rm MeV}/c$ is incompatible with the kinematics
of an undisturbed quasi-free spectator within the stated energy and angular
acceptance. In contrast, the spectator Jacobi momentum $k_{sF}$ naturally
lies on the $100$--$170~{\rm MeV}/c$ scale. Hence, the precise definition
of the momentum plotted in Ref.~\cite{Li} is essential.

There is, however, a second and more fundamental ambiguity. In the
$\alpha_0$ channel the measured final state is
\begin{equation}
{}^{12}{\rm C}({}^{16}{\rm O},\alpha_s\alpha_0){}^{20}{\rm Ne},
\end{equation}
and the two $\alpha$ particles are identical. Although an $\alpha$
detected within $\pm8^\circ$ was measured in coincidence with the
second $\alpha$, coincidence alone does not establish that the
forward particle is the original spectator $\alpha_s$. The decay
particle $\alpha_0$ from
\begin{equation}
{}^{12}{\rm C}+{}^{12}{\rm C}
\rightarrow\alpha_0+{}^{20}{\rm Ne}
\end{equation}
can also populate the forward angular acceptance. In addition,
post-decay rescattering of $\alpha_s$ by $\alpha_0$ or
${}^{20}{\rm Ne}$ can change the momentum of the original spectator.

Therefore, there are two distinct possibilities. If the forward
particle is an undisturbed $\alpha_s$, its intrinsic momentum is
kinematically restricted to $p_{sx}\lesssim80~{\rm MeV}/c$ and cannot
account for the full momentum interval shown in Ref.~\cite{Li}. If, on
the other hand, the forward particle is $\alpha_0$, or an $\alpha_s$
whose momentum has been modified by final-state rescattering, its measured
momentum no longer represents the initial intrinsic momentum
$\bm p_{sx}$. In either case, the measured forward-$\alpha$ distribution
cannot, without an independent demonstration of spectator identity and
negligible final-state rescattering, be identified with
$|\phi_{sx}(\bm p_{sx})|^2$ or used by itself as evidence for the
quasi-free mechanism.

\section{Plane-wave momentum distribution}
\label{sec:pw}

In Ref.~\cite{Li}, the momentum distribution of the spectator
$\alpha_s$ particle in ${}^{16}{\rm O}$ was calculated using a
Woods--Saxon potential with $R=4.65$ fm, $a=0.65$ fm, and
$V=32.45$ MeV and was compared with the experimental distribution
plotted as a function of the momentum denoted there by $p_s$. As
discussed in Sec.~\ref{sec:psx_psf}, the identification of this
experimental variable with the intrinsic momentum $p_{sx}$ is not
kinematically evident. Since the Coulomb
radius was not specified in Ref.~\cite{Li}, the present calculation
employs a Woods--Saxon potential with
$R_N=4.65$ fm, $R_C=3.15$ fm, $a=0.65$ fm, and $V=38.70$ MeV.

For the external part of the $s+x$ bound-state wave function, the
improved representation
\begin{equation}
\phi_{\rm imp}(r_{sx})\propto e^{-\kappa_{sx} r_{sx}}
\left(
\frac{1}{r_{sx}^2}
-\frac{c}{r_{sx}^3}
\right),
\qquad
r_{sx}\geq3~\mathrm{fm},
\label{eq:phiimp}
\end{equation}
is used. For the reaction considered here, the fitted value is
\[
c=3.44~\mathrm{fm}.
\]
The corresponding plane-wave momentum-space amplitude is
\begin{equation}
F_{\rm PW}(p_{sx})=4\pi\int_{3\,\mathrm{fm}}^{\infty}
{\rm d}r_{sx}\,r_{sx}^2\,j_0(p_{sx}r_{sx})\,\phi_{\rm imp}(r_{sx}).
\label{eq:pwtransform}
\end{equation}

\begin{figure}[t]
\centering
\includegraphics[width=\columnwidth]{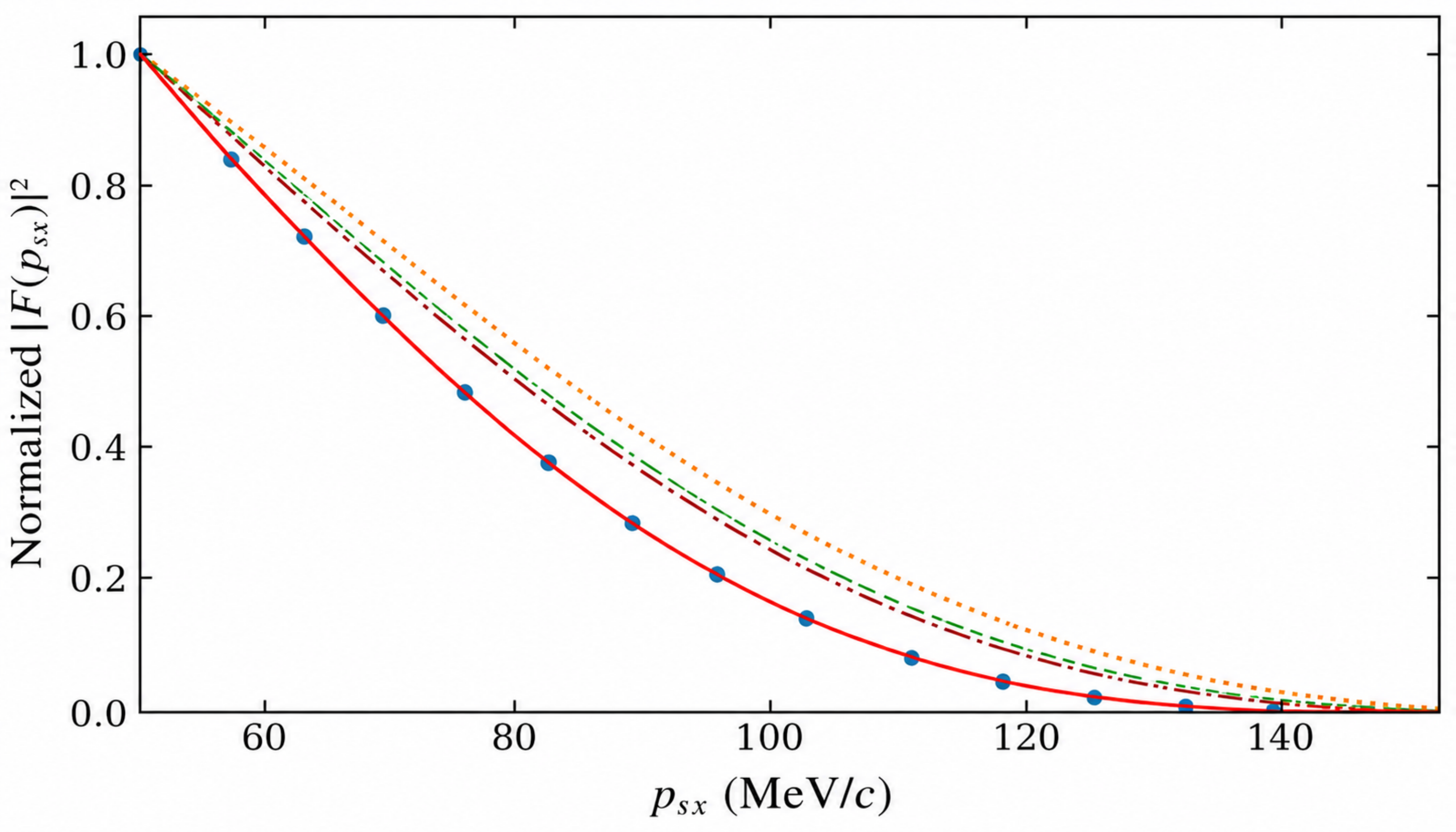}
\caption{Spectator momentum distribution. The blue dots denote the
reference distribution obtained from the Whittaker representation of
the asymptotic Woods–Saxon bound-state wave function with the potential
parameters specified in the text, while the curves show the corresponding
analytic approximations $\phi_{\rm imp}(r_{sx})$ given by
Eq.(\ref{eq:phiimp}) for different values of $c$. The best agreement
with the reference distribution is obtained for
$c=3.44~\mathrm{fm}$. All distributions are normalized at
$p_{sx}=50~\mathrm{MeV}/c$.}
\label{fig:pw_momentum}
\end{figure}

Figure~\ref{fig:pw_momentum} shows the spectator momentum distribution
for the bound $\alpha_s-{}^{12}{\rm C}$ configuration calculated using
Eq.(\ref{eq:phiimp}) and compares it with the reference momentum
distribution obtained from the asymptotic part of the Woods–Saxon
bound-state wave function. In the external region, this wave function
is represented by the corresponding Whittaker function. Thus, the
Whittaker momentum distribution shown in Fig.\ref{fig:pw_momentum}
provides the reference distribution associated with the Woods–Saxon
potential specified above. The value $c=3.44$ fm provides the best
agreement between the analytic representation of
Eq.~(\ref{eq:phiimp}) and this reference momentum distribution.

\section{Coulomb-distorted THM amplitude}
\label{sec:coulomb}

According to Ref.~\cite{muk2019}, the THM double-differential cross section
for the resonant subreaction
$x+A\to F^{*}\to b+B$ can be expressed in terms of the astrophysical
$S$ factor and the differential cross section; see Eq.~(35) of
Ref.~\cite{muk2019}. The latter contains the radial matrix element
${\cal M}^{\rm DWZR(prior)}$ given by Eq.~(30) of Ref.~\cite{muk2019}.
This DWBA-based matrix element determines the energy dependence of the
THM differential cross section.

The central objective of the present work is to derive an analytic
representation of ${\cal M}^{\rm DWZR(prior)}$ suitable for the forward
kinematic region. The numerical calculations of Ref.~\cite{muk2019},
performed with the FRESCO code~\cite{Thompson}, produced a backward peak
in the spectator angular distribution, characteristic of head-on and
near-head-on collisions dominated by low partial waves. In contrast, the
experimental acceptance in Ref.~\cite{Li}, as in Ref.~\cite{Tumino}, was
restricted to forward angles of the spectator $\alpha_s$. An exact
analytic representation is therefore required to determine the forward
behavior of the amplitude for THM reactions involving heavier charged
nuclei, for which Coulomb distortions cannot be neglected.

For simplicity, the superscript ${\rm DWZR(prior)}$ is omitted below.
The central quantity is then the zero-range prior-form transfer amplitude
\begin{align}
{\cal M}
=
&\int d\mathbf r_{sx}\,
\Psi_{-\mathbf k_{0}}^{C(+)}(\mathbf r_{sx})\,
\phi_{sx}(\mathbf r_{sx})\,
\Psi_{\mathbf k_{aA}}^{C(+)}
\left(
\frac{m_s}{m_a}\mathbf r_{sx}
\right).
\label{MDZRpr1}
\end{align}
For the final-state Coulomb-distorted wave, the representation
\begin{equation}
\Psi_{-\mathbf k_0}^{C(+)}
=
\Psi_{\mathbf k_0}^{C(-)*}
\end{equation}
is understood.

A truncated partial-wave expansion dominated by the lowest partial waves
cannot adequately describe the forward region relevant to the experiment.
The $\alpha_s$ acceptance in Ref.~\cite{Li} is restricted to approximately
$\pm8^\circ$ in the laboratory system. In this region, the amplitude
receives coherent contributions from a very large number of partial waves,
extending into the asymptotic large-$l$ domain. A direct numerical
partial-wave summation therefore becomes impractical and also obscures the
analytic structure responsible for the forward Coulomb behavior.

An analytic method is developed below that avoids the partial-wave
expansion and directly determines the forward amplitude. The resulting
expression is applicable to THM transfer reactions involving charged
nuclei and resonant states in the intermediate $x+A$ subsystem. It is
also applicable to the THM reaction of Ref.~\cite{Tumino}.

Such a DWBA description of the forward region is essential for assessing
the validity of the plane-wave approximation. In Refs.~\cite{Li,Tumino},
agreement between the measured spectator momentum distribution and that
generated from a Woods--Saxon bound-state wave function was used to support
the applicability of the PWA. Such agreement alone, however, does not
establish its validity. As shown below, the Coulomb-distorted THM amplitude ${\cal M}$
produces a spectator momentum distribution close to the Woods--Saxon
result while exhibiting an energy dependence that is qualitatively
different from that obtained in the PWA.

The derivation employs the Nordsieck integral~\cite{Nordsieck}, which
provides an exact analytic treatment of the convolution of the Yukawa
kernel
\begin{equation}
\frac{e^{-\kappa r}}{r}
\end{equation}
with the Coulomb-distorted waves in the initial and final channels.
Using the representation of the $s-x$ bound-state wave function introduced
in Eq.~(\ref{eq:phiimp}), the transfer amplitude becomes
\begin{align}
{\cal M}
&=
{\cal M}_{1}
-
{\cal M}_{2},
\label{eq:fullM}
\end{align}
where
\begin{align}
{\cal M}_{1}
&=
\int_{\kappa_{sx}}^{\infty}dy\,
\mathcal K(y),
\label{eq:dwzr}
\\[1ex]
{\cal M}_{2}
&=
3.44~{\rm fm}\,
\int_{\kappa_{sx}}^{\infty}dy
\int_{y}^{\infty}dt\,
\mathcal K(t).
\label{eq:dw2zr}
\end{align}
Here,
\begin{align}
\mathcal K(y)
&=
\int d\mathbf r_{sx}\,
\Psi_{-\mathbf k_{0}}^{C(+)}(\mathbf r_{sx})\,
\frac{e^{-y r_{sx}}}{r_{sx}}\,
\Psi_{\mathbf k_{aA}}^{C(+)}
\left(
\frac{m_s}{m_a}\mathbf r_{sx}
\right)
\label{Ky1}
\end{align}
is the Coulomb-distorted Yukawa amplitude.

The coordinate integral in Eq.~(\ref{Ky1}) is evaluated analytically
using the Nordsieck integral and can be expressed in terms of the Gauss
hypergeometric function ${}_{2}F_{1}$. This representation effectively
performs the coherent summation over the large angular momenta governing
the forward Coulomb amplitude and thereby eliminates the need for an
impractically large partial-wave expansion. The explicit expression in
terms of ${}_{2}F_{1}$ is not displayed because its lengthy form is not
required for the analysis below.

\section{Momentum-distribution test}
\label{sec:momentumtest}

\begin{figure}[t]
  \centering
  \includegraphics[width=\columnwidth]{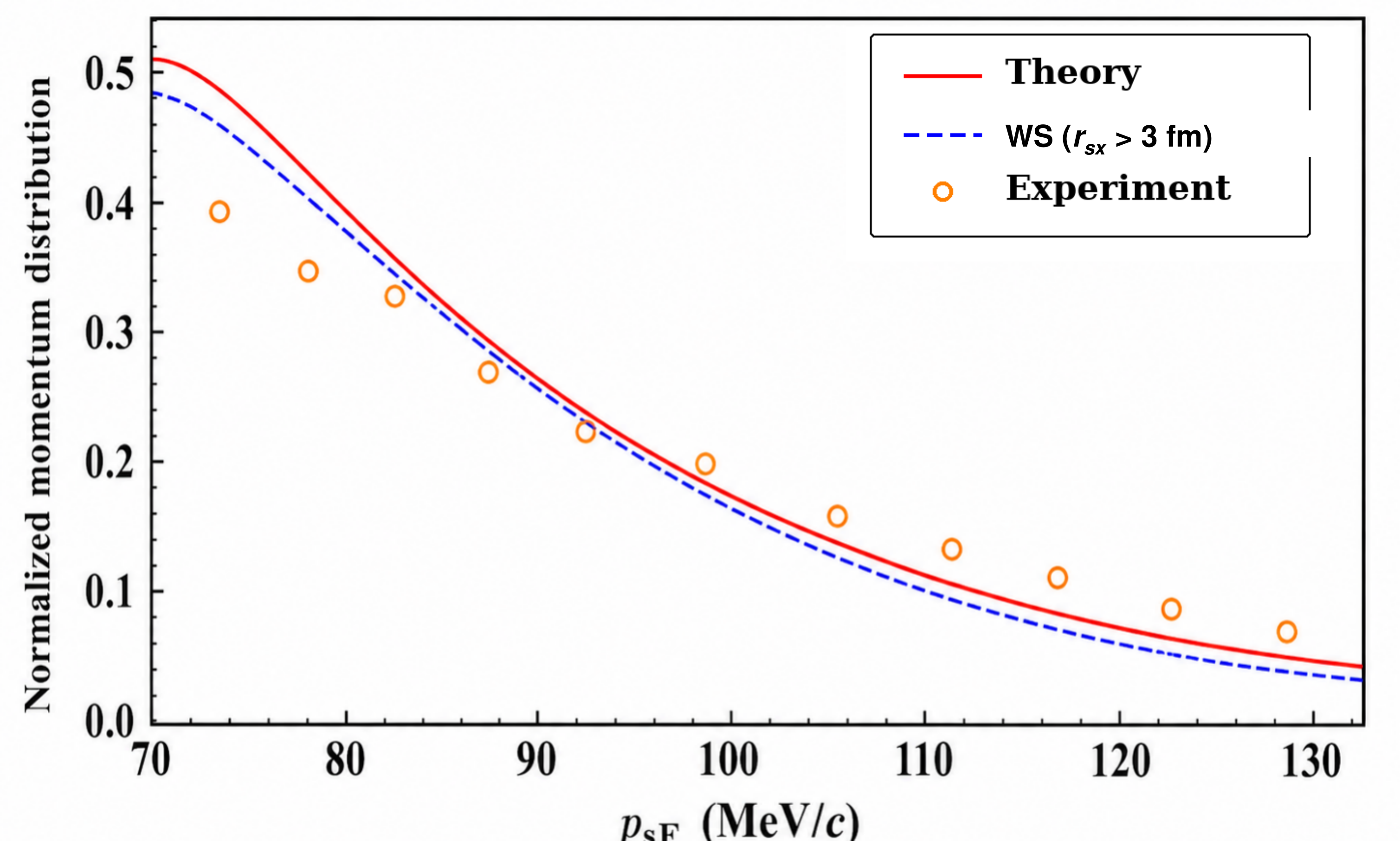}
  \caption{Normalized spectator momentum distribution. The open circles
  denote the experimental data from Ref.~\cite{Li}, plotted there versus
  the momentum variable $p_s$. The solid red curve shows the
  Coulomb-distorted calculation, and the dashed blue curve is the
  distribution generated by the Woods--Saxon (WS) bound-state wave
  function for $r_{sx}>3$ fm. The distinction between the experimental
  variable $p_s$, the spectator Jacobi momentum $k_{sF}$, and the
  intrinsic momentum $p_{sx}$ is discussed in Sec.~\ref{sec:psx_psf}.}
  \label{fig:momentumtest}
\end{figure}

Having derived an expression for the Coulomb-distorted amplitude, the
spectator momentum distribution can now be calculated with Coulomb
interactions included in both the initial and final states of the THM
reaction. In particular, the final-state Coulomb interaction experienced
by the spectator is treated explicitly. The resulting momentum distribution
is shown in Fig.~\ref{fig:momentumtest}.

The result shows that inclusion of the Coulomb distortions in the initial
and final states does not significantly modify the shape of the momentum
distribution generated by the Woods--Saxon bound-state wave function.
This observation is particularly important for assessing the validity of
the PWA. In Ref.~\cite{Li}, as in Ref.~\cite{Tumino}, agreement between
the calculated spectator momentum distribution and the experimental one was
used as an argument supporting the applicability of the PWA. The present
calculation demonstrates that this criterion is insufficient and therefore
cannot be used to justify the PWA in the analysis of these THM reactions.
Indeed, essentially the same spectator momentum distribution is obtained
when the strong Coulomb distortions in both the initial and final channels
are explicitly included. Therefore, reproduction of the experimental
momentum distribution cannot be regarded as evidence for the validity of
the PWA. The spectator momentum distribution is not a sensitive test of
the Coulomb distortions, although these distortions strongly affect the THM
reaction amplitude and, as shown below, its energy dependence.

The Woods--Saxon potential adopted in the present calculation differs from
that used in Ref.~\cite{Li}, whose parameters were adjusted to reproduce the
experimental momentum distribution. Consequently, the calculated distribution exhibits a small deviation from
the experimental data. This
difference originates from the adopted bound-state wave function rather than
from inclusion of the Coulomb distortions.

\section{Energy dependence of the THM amplitude}
\label{sec:energy}

Before considering the combined Coulomb-distorted result, it is useful to
separate the roles of the Coulomb interactions in the entrance and exit
channels. This separation also clarifies which part of the reaction mechanism
is primarily responsible for the strong energy dependence of the THM
amplitude.

\subsection{Impact of the initial-state Coulomb interaction}
\label{sec:initial_impact}

\begin{figure}[t]
  \centering
  \includegraphics[width=\columnwidth]{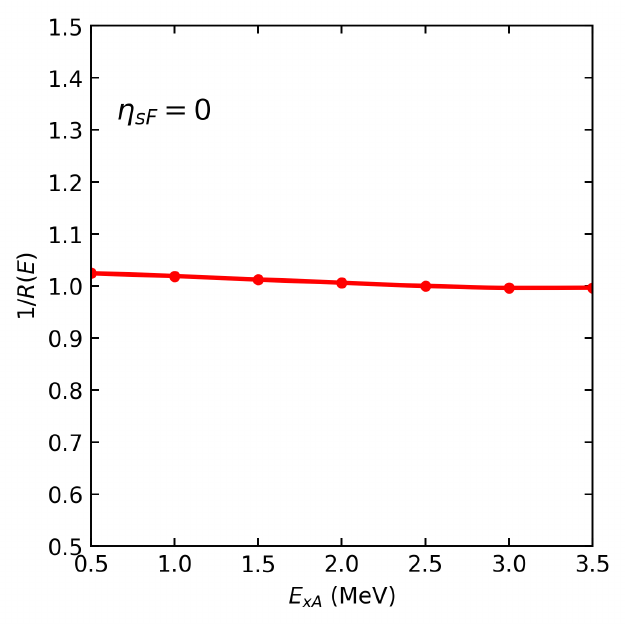}
  \caption{Inverse ratio $1/{\cal R}(E)$ calculated with the final-state
  Coulomb interaction switched off, $\eta_{sF}=0$, while retaining the
  initial-state Coulomb distortion. The ratio is normalized to unity at
  $E_{xA}=1.0$ MeV.}
  \label{fig:eta_sF_zero}
\end{figure}

To isolate the initial-state effect, the Coulomb distortion in the
${}^{16}{\rm O}+{}^{12}{\rm C}$ entrance channel is retained while the
final-state Coulomb interaction of the spectator is switched off. In
Eq.~(\ref{MDZRpr1}) this corresponds to replacing the final-state Coulomb
wave by the corresponding plane wave while keeping
$\Psi_{\mathbf k_{aA}}^{C(+)}$ unchanged. The entrance-channel Coulomb
interaction is strong, with $\eta_i=5.71$, and therefore substantially
modifies the magnitude of the transfer amplitude relative to the PWA.
However, because the beam energy and hence the entrance-channel Coulomb
parameter are fixed, this distortion by itself produces only a comparatively
weak dependence on the binary energy $E=E_{xA}$ over the interval of
interest. Thus the initial-state Coulomb interaction is important for the
absolute reaction amplitude but is not the principal source of the strong
energy dependence found below. This is illustrated in
Fig.~\ref{fig:eta_sF_zero}, where the final-state Coulomb interaction is
switched off by setting $\eta_{sF}=0$. The resulting inverse ratio remains
close to unity throughout the considered energy interval.

\subsection{Impact of the final-state Coulomb interaction}
\label{sec:final_impact}

\begin{figure}[t]
  \centering
  \includegraphics[width=\columnwidth]{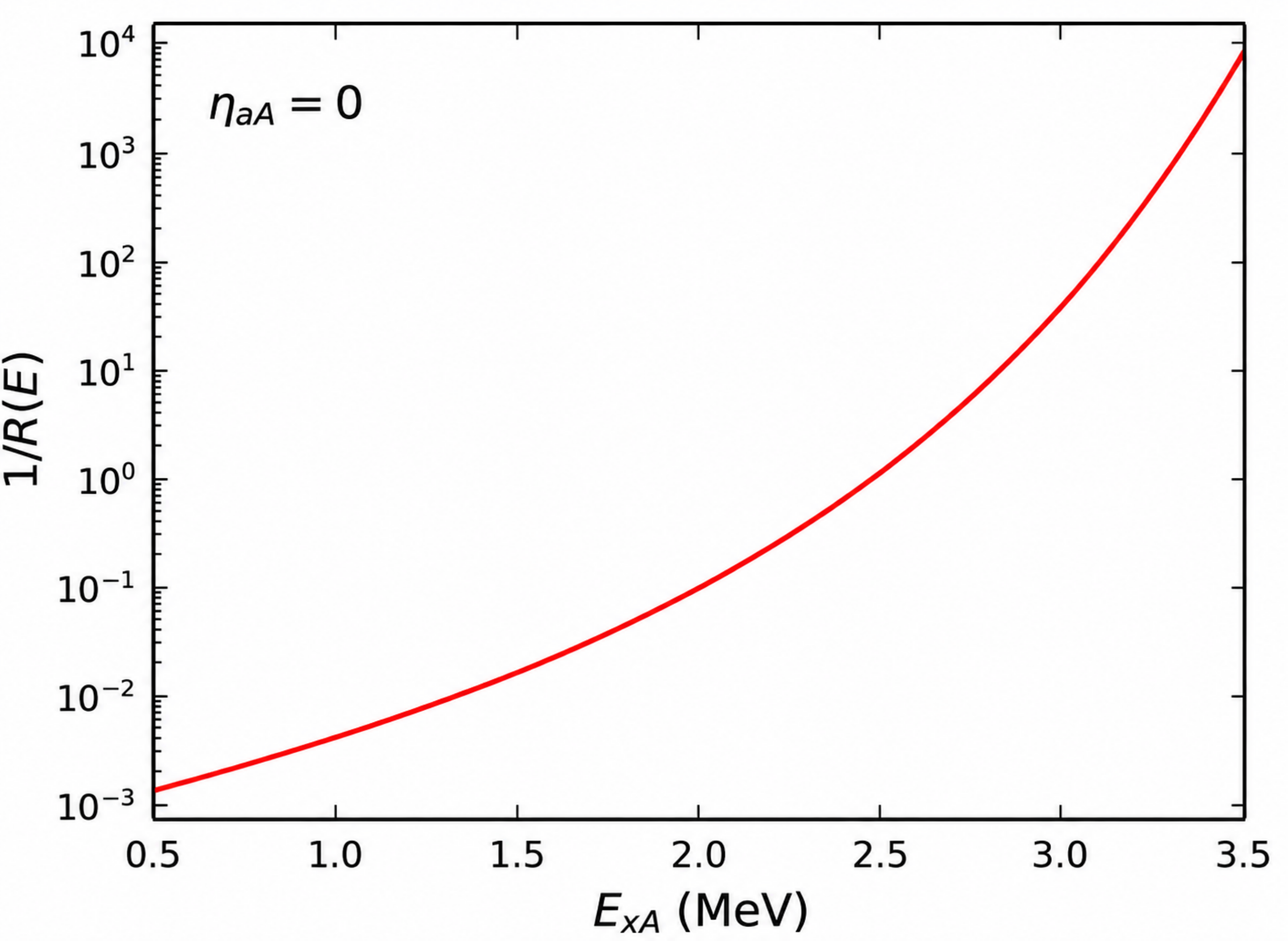}
  \caption{Inverse ratio $1/{\cal R}(E)$ calculated with the
  entrance-channel Coulomb interaction switched off, $\eta_{aA}=0$, while
  retaining the final-state Coulomb distortion. The ratio is normalized to unity at $E=2.5$ MeV. }
  \label{fig:eta_aA_zero}
\end{figure}

The situation changes qualitatively when the Coulomb interaction in the
final spectator channel is included. As $E$ increases, the energy available
for the relative motion of the spectator $\alpha_s$ and the intermediate
${}^{24}{\rm Mg}^{*}$ subsystem decreases. For example,
$E_{\alpha_s-{}^{24}{\rm Mg}^{*}}$ changes from $2.34$ MeV at
$E=2.5$ MeV to $1.34$ MeV at $E=3.5$ MeV. The corresponding final-state
Coulomb distortion therefore varies strongly with $E$. Calculations in
which the entrance-channel Coulomb distortion is suppressed while the
final-state Coulomb interaction is retained show that this interaction
provides the dominant energy-dependent contribution. Consequently, the
strong variation of the full Coulomb renormalization factor originates
primarily from the final-state interaction of the spectator with the
charged residual system. The initial-state distortion remains important
for the magnitude and interferes coherently with the final-state
contribution, so that both interactions must be retained in the complete
THM amplitude. The isolated final-state effect is shown in
Fig.~\ref{fig:eta_aA_zero}, obtained by suppressing the entrance-channel
Coulomb distortion, $\eta_{aA}=0$. In contrast to the $\eta_{sF}=0$ case,
$1/{\cal R}(E)$ changes by several orders of magnitude, demonstrating the
dominant role of the final-state Coulomb interaction in generating the
energy dependence.

\subsection{Combined Coulomb effect}
\label{sec:combined_impact}

The decisive difference between the plane-wave and Coulomb-distorted
calculations appears in their dependence on the relative energy
$E\equiv E_{xA}$. To quantify this difference, consider
\begin{equation}
{\cal R}(E)=
\frac{|{\cal M}(E)|^2}{|{\cal M}(E_{\rm norm})|^2},
\label{eq:ratio}
\end{equation}
where $E_{\rm norm}$ is the adopted normalization energy, so that
${\cal R}(E_{\rm norm})=1$.

\begin{figure}[t]
  \centering
  \includegraphics[width=\columnwidth]{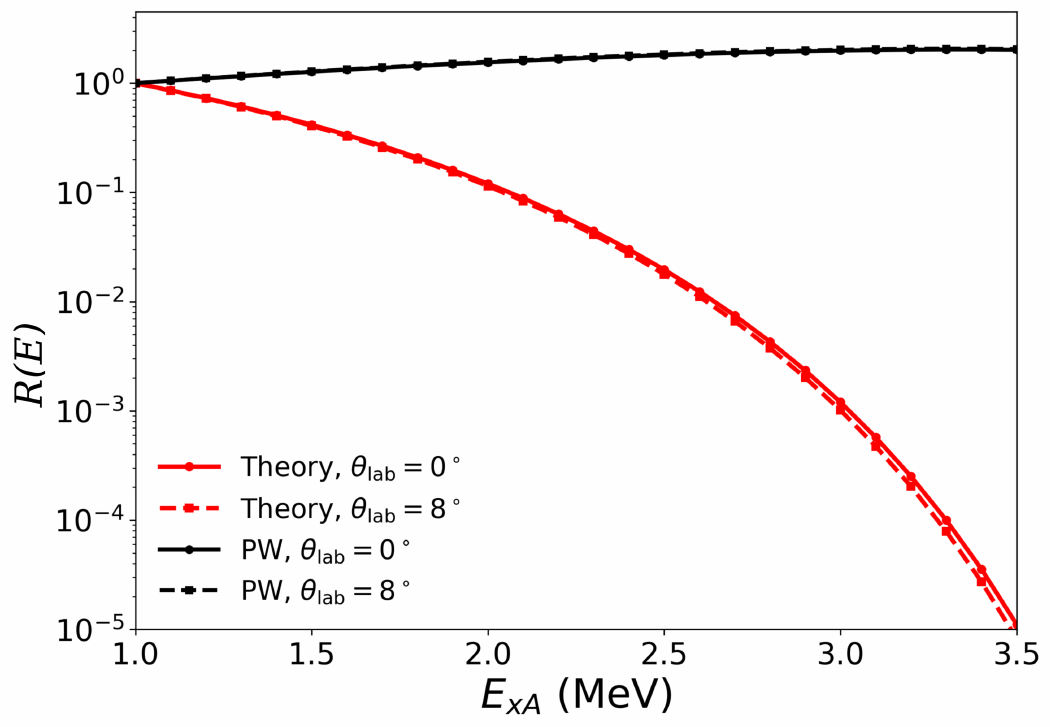}
  \caption{Energy dependence of the ratio
${\cal R}(E)=|{\cal M}(E)|^2/|{\cal M}(1.0~{\rm MeV})|^2$.
All curves are normalized to unity at $E=1.0$ MeV. The red curves are
obtained with Coulomb distortions included in both the initial and final
channels, whereas the black curves show the corresponding PWA results.
The solid curves correspond to $\theta_s^{\rm lab}=0^\circ$ and the dashed
curves to $\theta_s^{\rm lab}=8^\circ$.}
  \label{fig:energy_dependence}
\end{figure}

Calculations are performed for the spectator laboratory angles
$\theta_s^{\rm lab}=0^\circ$ and $8^\circ$. For the latter, the laboratory
angle is transformed to the corresponding energy-dependent c.m.\ angle.
The results are shown in Fig.~\ref{fig:energy_dependence}. In this figure,
it is convenient to adopt $E_{\rm norm}=1.0$ MeV.

The Coulomb-distorted amplitude decreases rapidly with increasing $E$.
Equivalently, the inverse ratio $1/{\cal R}(E)$ rises by approximately five orders of magnitude over the energy interval shown, whereas the corresponding
PWA ratio exhibits only a weak energy dependence. The very small difference between the
results at $\theta_s^{\rm lab}=0^\circ$ and $8^\circ$ demonstrates that
this qualitative behavior persists over the experimental angular
acceptance.

\begin{figure}[t]
 \centering
 \includegraphics[width=\columnwidth]{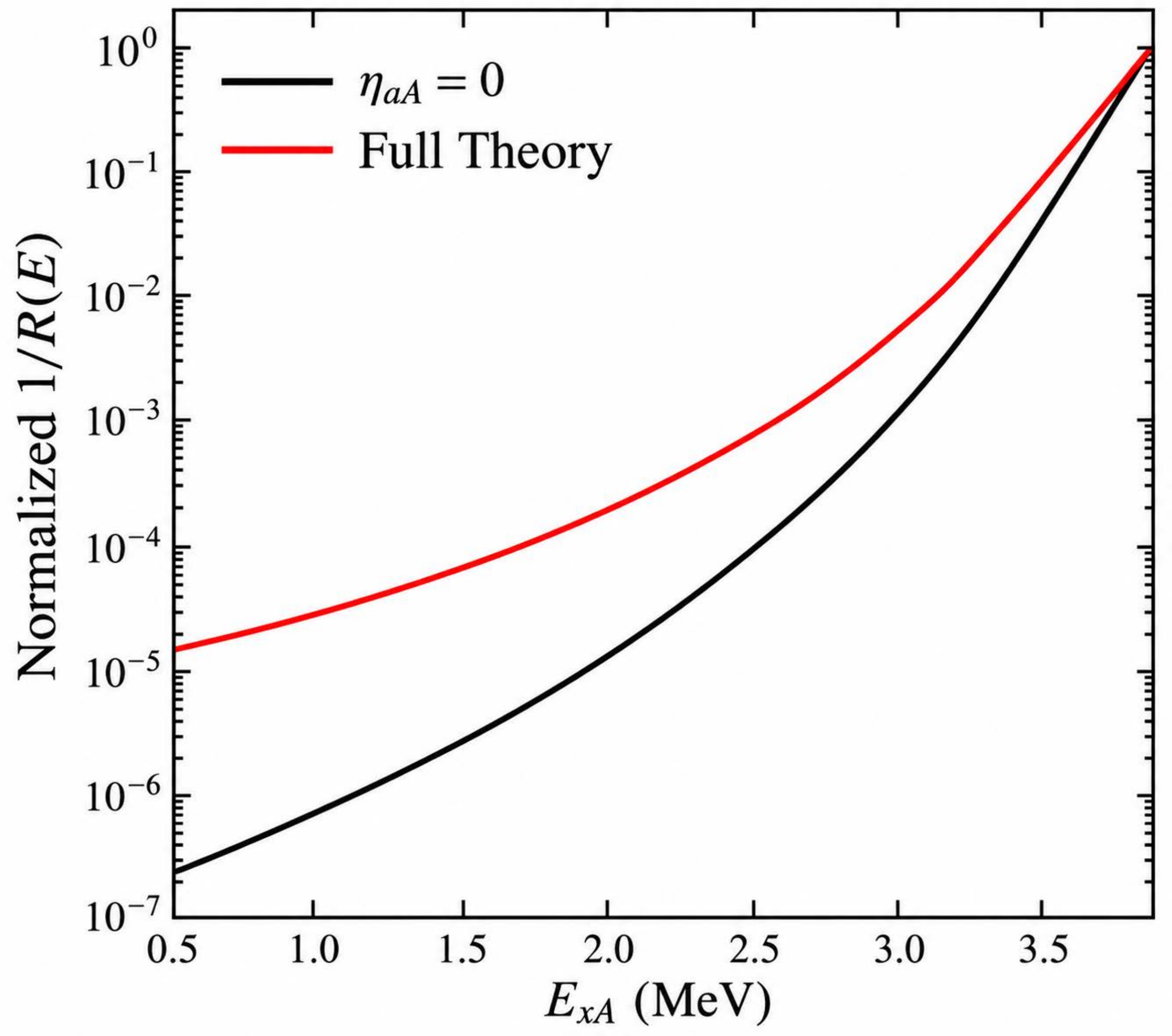}
 \caption{The red line is the inverse Coulomb renormalization factor
$1/{\cal R}(E)=|{\cal M}(3.5~{\rm MeV})|^2/|{\cal M}(E)|^2$
for the astrophysical $S$ factor extracted from the THM data of
Ref.~\cite{Li}. This renormalization factor is calculated taking into account 
the Coulomb interactions in the initial and final states.
For comparison, the inverse Coulomb renormalization
factor for $\eta_{aA}=0$ is also shown. Both lines are normalized
at $3.5$ MeV.}
\label{fig:SfctrCRen}
\end{figure}

According to Ref.~\cite{muk2019}, extraction of the astrophysical $S$
factor from the THM cross section requires division by the Coulomb
renormalization factor ${\cal R}(E)$. The corresponding factor
$1/{\cal R}(E)$ is shown in Fig.~\ref{fig:SfctrCRen}. Its strong energy
dependence provides
another compelling demonstration of the failure of the PWA for the
analysis of this THM reaction.

In Fig.~\ref{fig:SfctrCRen}, the inverse Coulomb renormalization factor
\begin{equation}
\frac{1}{{\cal R}(E)}=\frac{|{\cal M}(3.5~{\rm MeV})|^2}{|{\cal M}(E)|^2}
\end{equation}
is shown for the astrophysical $S$ factor extracted from the THM data of
Ref.~\cite{Li}. For comparison, the inverse Coulomb renormalization factor
calculated for $\eta_{aA}=0$, taken from Fig.~\ref{fig:eta_aA_zero}, is also
shown. It is seen that, while the initial-state Coulomb interaction alone
produces a practically flat Coulomb renormalization, its inclusion together
with the final-state Coulomb interaction significantly smooths the energy
dependence of the inverse Coulomb renormalization factor compared with that
obtained when only the final-state Coulomb interaction is included
($\eta_{aA}=0$).

\section{Updated Coulomb renormalization of the
$S(\alpha_0)$ factor from Ref.~\cite{Tumino}}
\label{sec:Tumino_update}
In Ref.~\cite{muk2019}, the Coulomb renormalization of the $S$ factor
extracted in Ref.~\cite{Tumino} was evaluated using a DWBA amplitude
describing the backward peak of the spectator angular distribution.
 The DWBA amplitude
was evaluated numerically with the FRESCO code~\cite{Thompson} using a
finite partial-wave expansion.

The present formulation provides a more general expression for the  THM
amplitude  that describes the forward peak
of the spectator angular distribution. This amplitude is more appropriate
for evaluating the Coulomb renormalization of the $S$ factor extracted in
Ref.~\cite{Tumino}, since the experimental kinematics selected the forward
spectator angular region.

\begin{figure}[t]
  \centering
  \includegraphics[width=\columnwidth]{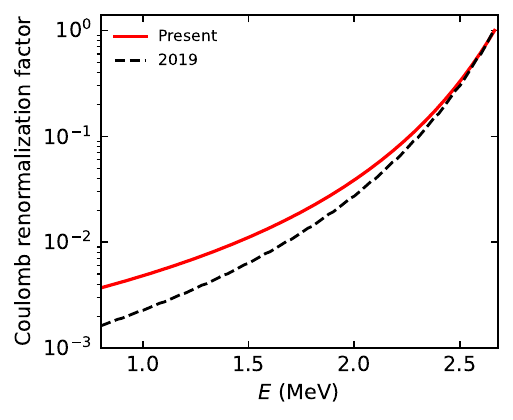}
  \caption{Comparison of the Coulomb renormalization factors for the
  $S(\alpha_0)$ factor extracted in Ref.~\cite{Tumino}. The dashed black curve
  shows the renormalization factor obtained in Ref.~\cite{muk2019},
  whereas the solid red curve shows the present forward-angle result.}
  \label{fig:THMFig7}
\end{figure}

A comparison of the Coulomb renormalization factor obtained in
Ref.~\cite{muk2019} with the present result for $1/{\cal R}(E)$ is shown
in Fig.~\ref{fig:THMFig7}. The difference between the two renormalization
factors reflects the different angular regions described by the
corresponding reaction amplitudes.

\begin{figure}[t]
  \centering
  \includegraphics[width=\columnwidth]{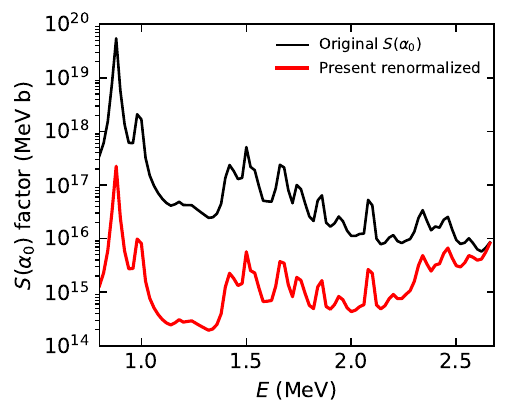}
  \caption{Comparison of the $S(\alpha_0)$ factors from
  Ref.~\cite{Tumino}. The black curve is the originally extracted
  $S(\alpha_0)$ factor, whereas the red curve shows the result obtained
  after applying the present Coulomb renormalization from Fig. \ref{fig:THMFig7}.}
  \label{fig:THMFig8}
\end{figure}

Figure~\ref{fig:THMFig8} compares the original
$S(\alpha_0)$ factor reported in Ref.~\cite{Tumino} with the result obtained
after applying the present Coulomb renormalization. The Coulomb correction
substantially modifies both the magnitude and the energy dependence of the
extracted $S$ factor, particularly toward the lower-energy region.

\section{Summary and conclusions}
\label{sec:conclusion}

The validity of the PWA used in the THM analysis of the
$^{12}\mathrm{C}+{}^{12}\mathrm{C}$ fusion reaction in Ref.~\cite{Li}
has been examined. The first principal result concerns the spectator
momentum distribution. Agreement between the measured $\alpha_s$ momentum
distribution and the momentum distribution of the $\alpha_s$ particle
inside ${}^{16}\mathrm{O}$ is not sufficient to establish the validity of
the plane-wave reaction mechanism. In fact, inclusion of the Coulomb
distortions in both the initial and final channels leaves the shape of the
spectator momentum distribution essentially unchanged. Consequently,
reproduction of the measured spectator momentum distribution cannot by
itself serve as a criterion for the applicability of the PWA to these THM
reactions.

The second principal result concerns the energy dependence of the extracted
astrophysical $S$ factor. Inclusion of the Coulomb distortions in the initial
and final channels produces a dependence on the
${}^{12}\mathrm{C}$–${}^{12}\mathrm{C}$ relative energy $E$ that differs
dramatically from that obtained in the PWA. After application of the Coulomb renormalization, the $S$ factor extracted from the THM data of Ref.~\cite{Li} decreases toward lower
energies by approximately five orders of magnitude over the energy
interval $0.5\leq E\leq 3.5$ MeV. Thus, rather than exhibiting the flat low-energy behavior
implied by the PWA analysis, the Coulomb-renormalized $S$ factor displays
a trend resembling hindrance-type behavior ~\cite{Jiang} or that predicted
by microscopic calculations ~\cite{Descouvemont}.

The updated Coulomb renormalization of the THM data of Ref. ~\cite{Tumino}
provides a further consequence of the present formalism. The forward-angle
amplitude ${\cal M}(E)$ yields a renormalization factor that differs
significantly at low energies from that obtained previously using the
backward-angle DWBA amplitude. Application of the updated renormalization
therefore strongly modifies both the magnitude and the energy dependence
of the $S(\alpha_0)$ factor extracted from the THM data of
Ref.~\cite{Tumino}.

An additional issue concerns the comparison of the two THM measurements.
Since the THM does not determine the absolute normalization of the
astrophysical $S$ factor, the result of Ref.~\cite{Li} was normalized to
the older direct data of Ref. ~\cite{Mazarakis} in the region
$E\gtrsim 2.5$ MeV.  These data lie significantly above the more recent
direct measurements in the corresponding energy region ~\cite{Tan,Nippert}. The adopted
normalization therefore affects the absolute magnitude of the extracted
$S$ factor, but it cannot account for its characteristic energy dependence.
In this respect, the two THM results themselves exhibit substantial
differences. Despite the qualitative agreement with Ref.~\cite{Tumino}
stated in Ref.~\cite{Li}, the $S$ factor reported in Ref.~\cite{Tumino}
rises sharply toward lower energies and exhibits a rich resonance
structure, whereas the result of Ref.~\cite{Li} has a considerably flatter
overall energy dependence and contains fewer and substantially broader
structures. The qualitative agreement between the two THM results is therefore
limited, particularly with respect to their energy dependence and
resonance structure.

The presented results demonstrate that the treatment of Coulomb distortions is
essential for extracting the energy dependence of the astrophysical
$S$ factor from THM measurements involving heavy charged particles.
In particular, agreement between calculated and measured spectator momentum
distributions does not provide a sufficient test of the plane-wave
approximation. A reaction-mechanism treatment that consistently includes
the Coulomb interactions in the initial and final channels is required
before conclusions concerning the low-energy behavior of the
$^{12}\mathrm{C}+{}^{12}\mathrm{C}$ astrophysical $S$ factor can be drawn
from THM data.

When the present work was ready for submission, a new THM measurement
of the proton channels,
$^{12}{\rm C}({}^{16}{\rm O},\alpha_s p_{0})
{}^{23}{\rm Na}$ and  $^{12}{\rm C}({}^{16}{\rm O},\alpha_s p_{1})
{}^{23}{\rm Na}^{*}$
was reported in Ref.~\cite{Lip}. The conclusions of the present analysis
concerning the importance of Coulomb distortions in the THM reaction
mechanism remain applicable to these channels. In particular,
Table~\ref{tab_finstat_p} shows the relevant kinematic conditions for
the charged-particle subsystems of the
$^{12}{\rm C}({}^{16}{\rm O},\alpha_s p_{0})
{}^{23}{\rm Na}$ reaction at $E=2.5$ MeV.

\begin{table*}[t]
\centering
\small
\renewcommand{\arraystretch}{1.35}
 \caption{Relative energies, Coulomb barriers, and Coulomb parameters
for the relevant charged-particle subsystems of the
$p_0+{}^{23}{\rm Na}$ branch of the
$^{12}{\rm C}({}^{16}{\rm O},\alpha_s p_{0})
{}^{23}{\rm Na}^{(*)}$ THM reaction at $E=2.5$ MeV.}
\begin{tabular}{|l|c|c|c|c|l|}
\hline
System & $\theta$ & $E_{\rm rel}$ (MeV) & $V_C$ (MeV) &
$\eta$ & Coulomb regime \\
\hline
$\alpha_s-p_0$ & $0^\circ$ & $1.36$ & $0.80$ & $0.24$ &
above barrier \\
\hline
$\alpha_s-p_0$ & $60^\circ$ & $2.77$ & $0.80$ & $0.17$ &
above barrier \\
\hline
$\alpha_s-{}^{23}{\rm Na}$ & $0^\circ$ & $2.88$ & $5.11$ &
$3.77$ & subbarrier \\
\hline
$\alpha_s-{}^{23}{\rm Na}$ & $60^\circ$ & $2.62$ & $5.11$ &
$3.95$ & more strongly subbarrier \\
\hline
$\alpha_s-{}^{24}{\rm Mg}^{*}$ & --- & $2.34$ & $5.52$ &
$4.58$ & deeply subbarrier \\
\hline
\end{tabular}
\label{tab_finstat_p}
\end{table*}

Although the $\alpha_s-p_0$ subsystem is above its Coulomb barrier,
the $\alpha_s-{}^{23}{\rm Na}$ subsystem remains subbarrier.
More importantly, the $\alpha_s-{}^{24}{\rm Mg}^{*}$ relative motion
is deeply subbarrier, with $E_{\rm rel}=2.34$ MeV compared with the
Coulomb barrier $V_C=5.52$ MeV and the Coulomb parameter $\eta=4.58$.
Thus, the dominant Coulomb effect associated with the
$\alpha_s-{}^{24}{\rm Mg}^{*}$ motion is common to the two THM
reactions. This Coulomb effect is established before the decay of the
intermediate ${}^{24}{\rm Mg}^{*}$ state and is therefore independent
of whether this state subsequently decays into
$\alpha_{0,1}+{}^{20}{\rm Ne}({\rm g.s.},\,{\rm exc.})$,
$p_0+{}^{23}{\rm Na}({\rm g.s.})$, or
$p_1+{}^{23}{\rm Na}^{*}$.
Consequently, the use of the proton decay channels does not remove
the principal Coulomb-distortion effect discussed in the present work.
Moreover, since the dominant Coulomb effect originates from the
$\alpha_s+{}^{24}{\rm Mg}^{*}$ relative motion prior to the decay of
${}^{24}{\rm Mg}^{*}$, the dominant Coulomb-renormalization mechanism
identified for the reactions leading to the final states
$\alpha_s+\alpha_{0,1}+{}^{20}{\rm Ne}({\rm g.s.},\,{\rm exc.})$
remains operative for the reactions leading to
$\alpha_s+p_{0,1}+{}^{23}{\rm Na}({\rm g.s.},\,{\rm exc.})$.


\clearpage

\begin{thebibliography}{99}
\bibitem{NanWang}
W. Nan, Y. Wang, J. Su, \textit{et al.},
Determination of resonances in Gamow window for
$^{12}\mathrm{C}+{}^{12}\mathrm{C}$ fusion reaction via thick-target
inverse kinematics method,
Phys. Lett. B \textbf{862}, 139341 (2025),
doi:10.1016/j.physletb.2025.139341.

\bibitem{Tumino}
A. Tumino, C. Spitaleri, M. La Cognata, \textit{et al.},
An increase in the $^{12}\mathrm{C}+{}^{12}\mathrm{C}$ fusion rate
from resonances at astrophysical energies,
Nature \textbf{557}, 687 (2018),
doi:10.1038/s41586-018-0149-4.

\bibitem{Li}
C. Li, H. Jia, Q. Wen, \textit{et al.},
$S^*(E)$ measurement of the
$^{12}\mathrm{C}({}^{12}\mathrm{C},\alpha){}^{20}\mathrm{Ne}$
reaction at astrophysical energies via the Trojan horse method
with $^{16}\mathrm{O}$ quasi-free breakup,
Phys. Lett. B \textbf{879}, 140675 (2026),
doi:10.1016/j.physletb.2026.140675.

\bibitem{TypelBaur}
S. Typel and G. Baur,
Theory of the Trojan--Horse method,
Ann. Phys. \textbf{305}, 228 (2003),
doi:10.1016/S0003-4916(03)00060-5.

\bibitem{muk2019}
A. M. Mukhamedzhanov, D. Y. Pang, and A. S. Kadyrov,
Astrophysical factors of
$^{12}\mathrm{C}+{}^{12}\mathrm{C}$ fusion extracted using the
Trojan horse method,
Phys. Rev. C \textbf{99}, 064618 (2019).

\bibitem{Thompson}
I. J. Thompson,
Coupled reaction channels calculations in nuclear physics,
Comput. Phys. Rep. \textbf{7}, 167 (1988),
doi:10.1016/0167-7977(88)90005-6.

\bibitem{Nordsieck}
A. Nordsieck,
Reduction of an integral in the theory of bremsstrahlung,
Phys. Rev. \textbf{93}, 785 (1954),
doi:10.1103/PhysRev.93.785.


\bibitem{Jiang}
C. L. Jiang, H. Esbensen, K. E. Rehm, B. B. Back, R. V.
F. Janssens, J. A. Caggiano, P. Collon, J. Greene, A. M.
Heinz, D. J. Henderson, I. Nishinaka, T. O. Pennington, and
D. Seweryniak, Unexpected behavior of heavy-ion fusion cross
sections at extreme sub-barrier energies, Phys. Rev. Lett. {\bf 89},
052701 (2002).

\bibitem{Descouvemont}  P. Descouvemont, Towards a microscopic description of fusion
at stellar energies, Phys. Rev. C {\bf 113}, 034613 (2026).

\bibitem{Mazarakis} M. G. Mazarakis and W. E. Stephens, Experimental measurements
of the 12C+12C nuclear reactions at low energies, Phys.
Rev. C {\bf 7}, 1280 (1973).

\bibitem{Tan}  W. P. Tan, A. Gula, K. Lee, A. Majumdar, S. Moylan, O. Olivas-
Gomez, Shahina, M. Wiescher, E. F. Aguilera, D. Lizcano,
E. Martinez-Quiroz, and J. C. Morales-Rivera, Coincident
measurement of the 12C+12C fusion cross section via the differential thick-target technique, 
Phys. Rev. C {\bf 110}, 035808  (2024).

\bibitem{Nippert}   J. Nippert et al. (STELLA Collaboration), Refining the deep
sub-barrier 12C+12C fusion excitation function with the
STELLA apparatus, Phys. Rev. C {\bf 111}, 065804 (2025).

\bibitem{Lip}
C. Li, H. Jia, Q. Wen, C. Lin, L. Yang, F. Yang, N. Ma,
T. Luo, X. Sun, S. Shao, and X. Wang, Indirect measurement of the $S^{*}(E)$ factor for
$^{12}\mathrm{C}({}^{12}\mathrm{C},p){}^{23}\mathrm{Na}$
at Gamow energies via the Trojan horse method with near-$0^\circ$
spectator detection, Phys. Rev. C \textbf{114}, 035801 (2026).



\end{thebibliography}
\end{document}